\documentclass[letterpaper,english,aps, nofootinbib, prd, twocolumn, lengthcheck, superscriptaddress]{revtex4-1}
\usepackage[T1]{fontenc}
\usepackage[latin9]{inputenc}
\usepackage{verbatim}
\usepackage{bm}
\usepackage{amsmath}
\usepackage{amssymb}
\usepackage{graphicx}
\usepackage{slashed}
\usepackage{wrapfig}
\usepackage{bbm}
\usepackage[caption=false]{subfig}
\usepackage{verbatim}

\pdfpagewidth\paperwidth

\usepackage[usenames, dvipsnames]{xcolor}
\definecolor{grayflag}{gray}{0.6}

\newcommand{\blueflag}[1]{{\color{blue} #1}}

\newcommand{\Tr}{\mbox{Tr}}
\newcommand \beq{\begin{eqnarray}}
\newcommand \eeq{\end{eqnarray}}
\newcommand {\EQCDs}{  E_{\rm QCD}^{\rm s}}
\newcommand {\EQCDv}{  E_{\rm QCD}^{\rm v}}
\newcommand {\QCD}{ {\rm QCD}}

\newcommand \as{\alpha_s}

\usepackage[colorlinks=true,linktocpage=true,linkcolor=blue,citecolor=blue]{hyperref}

\makeatother

\usepackage{babel}

\makeatletter

\begin{document}

\title{Effective repulsion in dense quark matter   from  non-perturbative gluon exchange}
 
\author{Yifan Song }
\address{Department of Physics, University of Illinois at Urbana-Champaign, 1110 W. Green Street, Urbana, Illinois 61801, USA\\}
\address{Interdisciplinary Theoretical and Mathematical Sciences (iTHEMS) Program, RIKEN, Wako, Saitama 351-0198, Japan}

\author{ Gordon Baym}
\address{Department of Physics, University of Illinois at Urbana-Champaign, 1110 W. Green Street, Urbana, Illinois 61801, USA\\}
\address{Interdisciplinary Theoretical and Mathematical Sciences (iTHEMS) Program, RIKEN, Wako, Saitama 351-0198, Japan}

\author{Tetsuo Hatsuda}
\address{Interdisciplinary Theoretical and Mathematical Sciences (iTHEMS) Program, RIKEN, Wako, Saitama 351-0198, Japan}
\address{Quantum Hadron Physics Laboratory, RIKEN Nishina Center, Wako, Saitama 351-0198, Japan}

\author{Toru Kojo}
\address{Key Laboratory of Quark and Lepton Physics (MOE) and Institute of Particle Physics,
Central China Normal University, Wuhan 430079, China}
\address{Interdisciplinary Theoretical and Mathematical Sciences (iTHEMS) Program, RIKEN, Wako, Saitama 351-0198, Japan}
 
\date{\today}

\begin{abstract}

   A moderately strong vector repulsion between quarks in dense quark matter is needed to explain how a quark core can support neutron stars heavier than two solar masses.  We study this repulsion, parametrized by a  four-fermion interaction with coupling $g_V$,  in terms of non-perturbative gluon exchange in QCD in the Landau gauge. Matching the energy of quark matter, $g_V n_q^2$ (where $n_q$ is the number density of quarks) with the quark exchange energy calculated in QCD with a gluon propagator parametrized by a finite gluon mass $m_g$ and a frozen coupling $\alpha_s$, at moderate quark densities, we find that gluon masses $m_g$ in the range  200 - 600 MeV and $\alpha_s$ = 2 - 4 lead to a $g_V$  consistent with neutron star phenomenology.  Estimating the  effects of quark masses and a color-flavor-locked (CFL) pairing gap, we  find that $g_V$ can be well approximated by a flavor-symmetric, decreasing function of density.  We briefly discuss similar matchings for the isovector repulsion and for the pairing attraction.  
\end{abstract}

\maketitle

\section{Introduction}

Quarks are active degrees of freedom in the deep interior of massive neutron stars.   For a comprehensive review of quark matter and the QCD phase diagram, see \cite{NSReview,Fukushima2010} and references therein.  In Refs.~\cite{Masuda1,Masuda2,phenoQCD},  we constructed a family of quark-hadron equations of state in which matter is described at densities up to about twice nuclear saturation density, $n_0 \approx 0.16$ baryons per fm$^3$ by interacting nucleons, and at higher densities, $n_B \gtrsim$ 5-10 $n_0$, by interacting quark matter with a highly constrained interpolation of the equation of state between the two regimes.  This equation of state describes neutron star properties quite consistent with recent LIGO inferences from the binary neutron star merger, GW170817 \cite{GW170817}. Version QHC18 of this equation of state at zero temperature is reviewed in \cite{NSReview}, and the latest version, QHC19, was recently made available~\cite{qhc19,compose}.
      
       We describe quark  matter in terms of a Nambu--Jona-Lasinio (NJL) model with point interactions in the scalar,  diquark, and vector-isoscalar channels, with a Lagrangian schematically of the form~\cite{pairinglit,buballa_review}
\beq
 {\cal L}_{\rm int} =  G(\bar q q)^2 + H(\bar q\bar q)(qq) - g_V (\bar q \gamma^\mu q)^2,
   \label{NJL}
\eeq
where the vector repulsion in the isoscalar channel \cite{kunihiro} is needed for quark matter to support heavy neutron stars.  The resultant energy density from the vector repulsion  is $g_V n_q^2$, where $n_q=3n_B$ is the quark number density. 

While the scalar coupling $G$ and the ultraviolet cutoff $\Lambda_{\text{\tiny NJL}} $ of the NJL model
can be directly related to physical observables  such as the properties of pseudoscalar mesons, the vector repulsion at present is constrained only by comparing the equation of state of matter with observations of neutron stars.   As we have found in our QHC19 equation of state, to support neutron stars of masses above two solar masses (including the recently measured neutron star mass,  $2.17\pm 0.1$ solar masses in the pulsar PSR J0740+6620 \cite{cromartie}) requires that $g_V$ be well in the range 0.6-1.3 $G_0$, and $H$ in the range 1.35-1.65 $G_0$ \cite{qhc19},
where $G_0 = 1.835 \Lambda_{\text{\tiny NJL}}^{-2}$ with $\Lambda_{\text{\tiny NJL}}$ = 631.4, is the 
scalar coupling in the vacuum obtained by a fitting of pion observables  \cite{pairinglit,buballa_review}.
 Our aim in this paper is to explore further understanding  the structure of  Eq.~(\ref{NJL})
  in terms of QCD, and the strength of the vector repulsion 
in particular. This in turn improves the consistency of the NJL description with perturbative QCD at densities $n_B \gtrsim$ 50-100 $n_0$ \cite{Kurkela:2009gj,Freedman:1976ub}.   A simple Fierz transformation of the color-current -- color-current interaction, $\sim(\bar{q} \gamma_\mu \lambda^\alpha  q)^2$, leads to NJL couplings (\ref{NJL}) with the ratios $g_{V0}/G_0$ = 1/2  and $H_0/G_0$ = 3/4 (see Appendix A) \cite{buballa_review}
where the ``0'' continues to indicate vacuum values.  But in the fully interacting system, these ratios need not hold; as in QHC18 and QHC19 we focus on more general in-medium values of $g_V$ and $H$, studying here the density dependence of $g_V$ in particular.

   Since $g_V$ has dimensions of $\mbox{mass}^{-2}$, at asymptotically large densities, where the only energy scale is the quark Fermi momentum $p_F$,  $g_V$  should behave as $\sim \alpha_s /p_F^{2}$, where $\as$ is the QCD running coupling constant.
  On the other hand, in the highly non-perturbative vacuum at zero baryon density, the relevant scale is $\Lambda_{\rm QCD}$, 
  and we expect $g_V\sim \alpha_s / \Lambda_{\rm QCD}^{2}$.  Thus, the matter density dependence of $g_V$ can be ignored only when $p_F \ll  \Lambda_{\rm QCD}$, provided that $\as$ also freezes at low energy \cite{running}.   To smoothly connect $g_V$ at low density with that at high density, we adopt a model of massive gluons  \cite{Cornwall1982,Aguilar2016}  
  which includes non-perturbative generation of the gluon mass $m_g$ as well as the freezing of $\as$ in the Landau gauge  at low energies. 
  As we estimate, a gluon mass $m_g \sim 0.4$ GeV, and a moderately strong quark-gluon coupling $\as \sim 3$ at $5n_0$ (or similar values, shown in Fig.~\ref{asmg} below, with $\as/m_g^2$ roughly constant) can produce a strong enough $g_V\sim G_0$ to allow quark matter to support two-solar mass neutron stars.

At high density, where the matter tends to have equal population of up-, down-, and strange-quarks, flavor-singlet channels are much more important than non-singlet flavor channels.  This allows us to focus on the flavor-singlet scalar and vector couplings as well as CFL-type diquark pairing
\cite{alford2008}, favored for equal flavor population.  Flavor non-singlet interactions are nonetheless important at low densities (see Appendix B).

    This paper is organized as follows. In Sec.~\ref{free}, we present the single gluon exchange energy calculation starting with free quark and gluon Green's functions, at first using the two-loop running coupling constant in perturbative QCD.    The Landau pole in the running coupling leads to a strongly divergent result at a density $\lesssim 5n_0$.  
To avoid such a divergence, we consider, in Sec.~\ref{alphas}, a range of $\as$ and gluon masses, $m_g$, as estimated non-perturbatively below the one GeV scale, and comment on the connection to the QHC19 neutron star equation of state, constrained by neutron star observations, to sub-GeV theories of $\as$ and massive gluons.  
We also provide an approximate density-dependent parametrization of $g_V$ connecting the low density and high density limits.
Next in Sec.~\ref{chiral} we estimate effects on $g_V$ of a finite quark mass, $M_q$, arising from chiral condensation in the quark sector, and in  Sec.~\ref{pair} effects of diquark pairing.  As we show, a quark mass term tends to enhance $g_V$, while diquark pairing decreases it; both effects are suppressed by a gluon mass, and as a result a flavor-independent $g_V$ is a good approximation in the NJL model.  We summarize our discussion in Sec.~\ref{conclusion}.
In Appendix \ref{sec:fierz}, we show how the color current-current interactions can be rearranged via the Fierz transformation. 
In Appendix \ref{sec:nucleon},  we consider effective vector-isovector couplings, possibly important at intermediate and low
densities, and in Appendix \ref{sec:Ndelta}, we estimate the value of $H$ from the $N$-$\Delta$ mass splitting.

Throughout we work in natural units $\hbar = c =1$ with the metric $g_{\mu \nu} = {\rm diag}(1,-1,-1,-1)$, 
and focus on zero temperature with $N_f = N_c = 3$ and equal quark masses,  unless stated otherwise. 
We use the notation $\int_p = \int d^4 p/(2\pi)^4$.


\section{Weak coupling limit \label{free}}

   The quark-gluon interaction to leading order in $\alpha_s$ leads to the energy-density shift of the quark matter
\beq
   E_{\rm QCD} & = & -\frac{ i \pi \as}{2} \int d^4x \, \langle J_\mu^\alpha(x) A^\mu_\alpha(x) J_\nu^\beta(0)  A^\nu_\beta(0)\rangle, 
\label{Eex}
\eeq
where the expectation value is in a Fermi gas, $x=(t, \bm{x})$, and $t$ is integrated from 0 to $-i/T$ (with $T$ the temperature).
  The currents are $J_\mu^\alpha(x)  \equiv  \bar{q} (x) \gamma_\mu \lambda^\alpha q (x)$, where the $\lambda_\alpha$ are the color SU(3) Gell-Mann matrices normalized to ${\rm tr} \lambda_\alpha \lambda_\beta = 2\delta_{\alpha \beta}$.  

In the weak coupling limit, neglecting diquark pairing, Eq.~(\ref{Eex}) becomes the Fock term in terms of the two-quark interaction
\beq
 E_{\rm QCD} & \approx & \frac{\pi\alpha_s}{2}\int_{p,p'} \Tr \,\left[ S(p) \lambda_\alpha \gamma^\mu S(p') \lambda_\beta \gamma^\nu \right] D_{\mu \nu}^{\alpha \beta}(p-p') .  \nonumber \\
\label{Enopair}
\eeq
Here
 the trace $\Tr$ runs over flavor, color, and Dirac indices, and the integrations over frequencies $p_0$ and $p'_0$ are understood as the fermion Matsubara frequency summations, $\int dp_0 f(p_0) \to 2\pi i T\sum_n f(i\omega_n)$, where $\omega_n = 2\pi T n$, with $n = \pm 1/2, \pm 3/2,\,\dots$. The time-ordered quark Green's function is
\beq
S^{ab}_{ij}(x-y) &=& -i \langle \mathcal{T} q^a_i (x) \bar{q}^b_j (y)\rangle 
\eeq
and are denoted by $S(p)$ in momentum space; here $a,b$ are color indices and $i,j$ flavor indices.  The gluon Green's function is 
\beq
D^{\alpha \beta}_{\mu \nu}(x-y) = -i \langle \mathcal{T} A^\alpha_\mu (x) A^\beta_\nu (y)\rangle.
\eeq
With no medium modification of the gluons, $D$ in the Landau gauge takes the form in the momentum space,
\beq
D^{\alpha \beta}_{\mu \nu}(q) = -\delta^{\alpha \beta} \left( g_{\mu \nu} - \frac{q_\mu q_\nu}{q^2} \right) D(q).
\eeq
The full calculation of the energy leads to divergent Dirac sea contributions involving antiparticles.    Only the $g_{\mu \nu}$ term in $D^{\alpha \beta}_{\mu \nu}(q)$ contributes to the particle-particle exchange (Fock) energy, and we keep only this term.

The traces in Eqs.~(\ref{Enopair}) can be re-organized, via a Fierz transformation (see Appendix \ref{sec:fierz}), into traces over quark Green's functions in the quark-antiquark channels.  The NJL model contains two such channels: the scalar $\bar qq$ channel -- which is used to characterize the spontaneous chiral symmetry breaking -- and the vector-isoscalar $\bar q\gamma^\mu q$ channel.  The energies corresponding to the scalar and vector channels, after the Fierz expansion of Eqs.~(\ref{Enopair}),  denoted by $\EQCDs$ and $\EQCDv$, are
\beq
 E_{\rm QCD}^{\rm s} &=& -\frac{8\pi \as}{27} \int_{p,p'} \Tr  S(p) \,\Tr S(p') D(p-p'), 
 \label{Egs} \\
 E_{\rm QCD}^{\rm v}&=&  \frac{4\pi \as}{27} \int_{p,p'} \Tr [S(p)\gamma^\mu ] \Tr [S(p')\gamma_\mu ] D(p-p').  \nonumber \\
\label{Egv}
\eeq

     We first outline how these results are related to the effective $G$ and $g_V$ in the NJL model.  Since the detailed relation depends on the gluon propagator, we first illustrate the results in the two limiting extremes, low and high density.   
Owing to the non-perturbative infrared cutoff of order $\Lambda_{\QCD}$,
 the gluon propagator has a finite limit $D(q\rightarrow 0)$ at low energy;  
  thus at low densities we have  
\beq
 E_{\rm QCD}^{\rm s,v} &=& C_{\rm s,v} \alpha_s D(0) \left( \int_{p} \Tr \left[ S(p) \Gamma_{ {\rm s,v} } \right] \right)^2 \,,
 \label{Egv}
\eeq
where $C_{{\rm s}} = -8\pi/27 = - 2 C_{{\rm v}}$ and $\Gamma_{\rm s} =1$, and  provided that $\int_p \Tr[S (p) \gamma_j] = 0$, 
$\Gamma_{\rm v} = \gamma^0$.  In this form one can readily identify the NJL couplings as $G = 2g_V = C_{\rm s} \alpha_s D(0)$. 

  At higher densities we must keep the momentum dependence of the gluon propagators. 
For example, with massless free quark and gluon propagators,
\begin{align}
S^{0,ab}_{ij}(p) & =  \delta_{ab}\delta_{ij} \frac{\gamma_\mu p^\mu}{(p_0+\mu)^2 - {\bm p}^2}, \\
D^0(p) & = \frac{1}{p^2},
\end{align}
where $\mu$ is the quark chemical potential, we find the perturbative result,\footnote{While the full trace in Eq.~(\ref{Enopair}) contains contributions from both particles and antiparticles, we focus only on modifications due to non-zero particle densities here.} 
\begin{align}
E_{\rm QCD}^{\rm v}& = 24\pi \alpha_s \left( \int \frac{d^3 {p}}{(2\pi)^3} \frac{f(|{\bm p}|-\mu)}{|{\bm p}|} \right)^2 \,,
\label{eq:Eex_pQCD}
\end{align}
where $f(z) = [\mbox{exp}(z/T)+1]^{-1}$ is the Fermi distribution function;  at zero temperature (\ref{eq:Eex_pQCD}) reduces to
\begin{align}
E_{\rm QCD}^{\rm v}= \frac{3\alpha_s p_F^4}{2\pi^3}.
\label{eq:Eex_pQCD1}
\end{align}
This result is identical to the exchange energy of a highly relativistic electron gas to within flavor and color factors.\footnote{ 
 Equation (\ref{eq:Eex_pQCD}) includes the interactions between quark number densities $\bar{q} \gamma_0 q$, as well as those between spatial currents, $\bar{q} \gamma_j q$.    These contributions yield the matrix element, for on-shell momenta,
\beq
&& \Tr[S(p) \gamma^\mu] \, \Tr[S(p') \gamma_\mu]~
 \propto  \frac{ |\bm{p}| |\bm{p}'| - \bm{p}\cdot \bm{p'} }{ 2 |\bm{p}| |\bm{p}'| } \,,
\eeq
whose numerator cancels the pole from the massless gluon propagator, giving Eq.(\ref{eq:Eex_pQCD}).}$^,$\footnote{
  In deriving $E_{\rm QCD}^{\rm v}$ in Eq.~(12) from Eq.~(\ref{Egv}) with a momentum-dependent gluon propagator, the correlation functions $\langle \bar q \vec{\gamma} q\rangle$ are as important as $\langle \bar q \gamma^0 q\rangle$; the former is not included in  the NJL mean field description.  Such deficiency in the NJL model can be compensated by absorbing the contribution from  $\langle \bar q \vec{\gamma} q\rangle$ into the density dependence of $g_V$ itself; in this way, we can directly compare the NJL $g_V$ with the current definition of $g_V$ in terms of QCD parameters.}
 
   The vector repulsion contributes an energy density in the NJL model \cite{qhc19} 
 \begin{align}  
E_{\rm NJL}^{\rm v} = g_V n_q^2,
\label{eq:NJL-V}
\end{align}
which we identify with $E_{\rm QCD}^{\rm v}$ 
in the matching density region $\sim$ 5-20 $n_0$ corresponding to $p_F \sim$ 0.4-0.6 GeV, one finds 
\begin{align}
  g_V = \frac{\pi \alpha_s} {6p_F^2}.
\label{eq:gV_pQCD}
\end{align}

\begin{figure}
\includegraphics[scale=0.35]{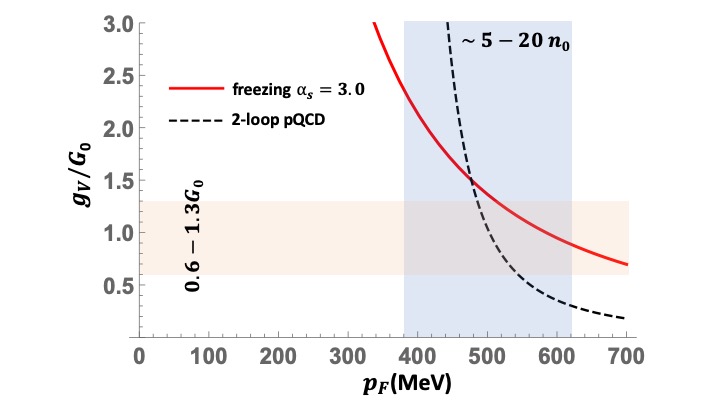}
\caption{\label{gV_pQCD}  The dashed line indicates the single gluon exchange result for $g_V$ in perturbative QCD as a function of the quark matter Fermi momentum, $p_F$.  
The horizontal shaded region shows the range of $g_V$ in QHC19 \cite{qhc19}, while the vertical shaded region shows the baryon density $\sim 5$-$20 n_0$.   The solid line indicates the result for $\alpha_s$ frozen at 3.0 
 at low energies \cite{running}.}
\end{figure}

  The solid line in Fig.~\ref{gV_pQCD} shows $g_V$ obtained using (\ref{eq:gV_pQCD}) and the
two-loop running coupling constant  $\alpha_s(\mu_q)$:
  \beq 
\alpha_{s}(\mu_q)=\frac{4\pi}{9\ln \tilde\mu^2}\left(1-\frac{64\ln{\ln \tilde\mu^2}}{81\ln \tilde\mu^{2}}\right),
 \eeq
with $ \tilde\mu \equiv  \mu_q/\Lambda_{{\rm QCD}}$ and  $\Lambda_{\rm QCD}$ = 340 MeV \cite{running}.
 The shaded horizontal band indicates  the range of (constant) $g_V$ 
in QHC19 \cite{qhc19}.  
Although  $g_V$ in Fig.~\ref{gV_pQCD} approaches the needed range below $20n_0$, the factor $p_F^{-2}$ and the running $\as$ near the Landau pole at $\Lambda_{\rm QCD}$ already causes strongly divergent behavior of $g_V$ even at $5n_0$ (corresponding to $p_F \sim 400$ MeV), in contrast to the simple treatment in NJL of $g_V$ as constant in this regime.  However, extending the pQCD calculation down to $\Lambda_{\rm QCD}$ is not reliable.
The solid line in Fig.~\ref{gV_pQCD}  shows $g_V$ for $\alpha_s$ frozen at 3.0 at low energies \cite{running}.
 Although the divergence from the Landau pole is removed in this case, $g_V$  still increases rapidly
at low energy.

 \section{Non-perturbative \boldmath$\as$ and massive gluons below one $\text{GeV}$ \label{alphas}}

  We now examine the consequences of the non-perturbative behavior of the strong coupling constant $\as$ and the gluon propagator below the 1 GeV scale.  For reviews,  see Refs.~\cite{running,Aguilar2016} and references therein.
In various non-perturbative approaches for the gluon sector  (lattice gauge theory, Schwinger-Dyson equations, and gauge/gravity duality) under  gauge fixing, $\as$ is of order unity below one GeV  (with freezing or decoupling behaviors in the deep infrared limit, \blueflag{$q \to 0$}).  Here we focus on gluons dynamically acquiring a mass, favored by the lattice results (and corresponding to the decoupling solution of the gluon Schwinger-Dyson equations in the Landau gauge),
\beq
D(p)  = \frac{1}{p^2 - m_g^2}.
\label{eq:gluon}
\eeq
Estimates of $m_g$ tend to lie in the range $\sim 500\pm200$ MeV \cite{Cornwall1982,Aguilar2016}.
   
   Equation~(\ref{eq:gluon}) regulates the divergent behavior of $g_V$ as $p_F\to 0$ in Fig.~\ref{gV_pQCD}  and  leads to
\beq
E_{\rm QCD}^{\rm v} (m_g) =E_{\rm QCD}^{\rm v}( 0) + \delta E_{\rm QCD}^{\rm v}(m_g),
\label{eq:exg}
\eeq
where (as in derivation of Eq.~(\ref{eq:Eex_pQCD}), $E_{\rm QCD}^{\rm v}(0)$ results from a cancellation between the massive gluon propagator with a part of quark matrix elements, while the remaining terms are proportional to $m_g^2$)
\beq
\delta E_{\rm QCD}^{\rm v}(m_g) & = & -\frac{3\alpha_s m_g^2}{2\pi^3} \int_0^{p_F} \int_0^{p_F}dp\,dp'\, \ln \left( 1+\frac{4pp'}{m_g^2} \right) \nonumber \\
 & = & \frac{3\alpha_s m_g^4}{8\pi^3} K(x),
 \label{eq17}
\eeq
where $z\equiv (2p_F/m_g)^2$ and
$K(z) \equiv 2z - (1+z)\ln (1+z) + \mbox{Li}_2(-z)$ with $\mbox{Li}_2(-z) \equiv \sum_{\ell=1}^\infty (-z)^\ell/ \ell^2$ the polylogarithm function with $n=2$.  
Thus one finds,
\beq
E_{\rm QCD}^{\rm v}(m_g) =  \frac{3\alpha_s p_F^4}{2\pi^3} \left( 1+ \frac{K(z)}{z^2} \right).
\label{eq:total-V}
\eeq
Note that for positive $z$,   $0 \le 1 + K(z)/z^2 < 1$, implying  that  the finite gluon mass softens the repulsion while
 keeping the total vector energy positive.
 
  Matching  Eq.~(\ref{eq:NJL-V}) with Eqs.~(\ref{eq:gV_pQCD}) and (\ref{eq:total-V}) one finds
\beq
g_V(p_F; z\gg 1) ~& \rightarrow &~  \frac{\, \pi \as \,}{6 p_F^2}, 
\nonumber \\
g_V(p_F; z\ll 1) ~& \rightarrow &~  \frac{\, 4\pi \as \,}{27m_g^2}.
\label{eq: g-V-limits}
\eeq

\begin{figure}
\includegraphics[scale=0.35]{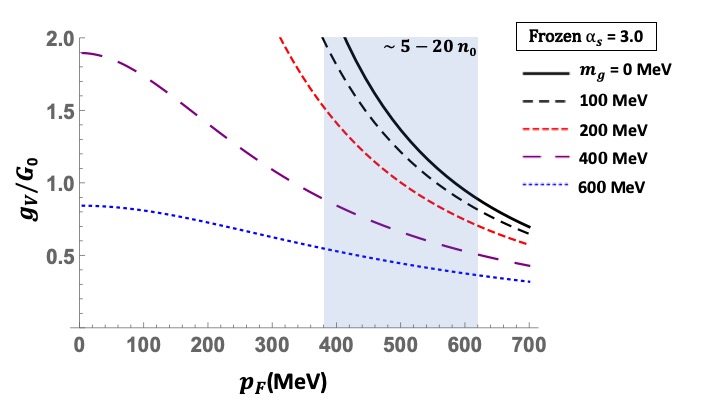}
\caption{\label{asmgpf}  (Color online) The vector coefficient $g_V$ as a function of quark Fermi momentum generated by a frozen $\as$ =3 below 1 GeV and different gluon masses $m_g$. } 
\end{figure}

 Figure~\ref{asmgpf} shows  $g_V$ for different gluon masses $m_g$ with a typical value of the frozen $\as = 3.0$ at low energies $\lesssim 1$ GeV \cite{running}. 
In the infrared $g_V$ is regulated by the gluon mass, $m_g$, so that there is no divergent behavior at $p_F=0$.

\begin{figure}
\includegraphics[scale=0.35]{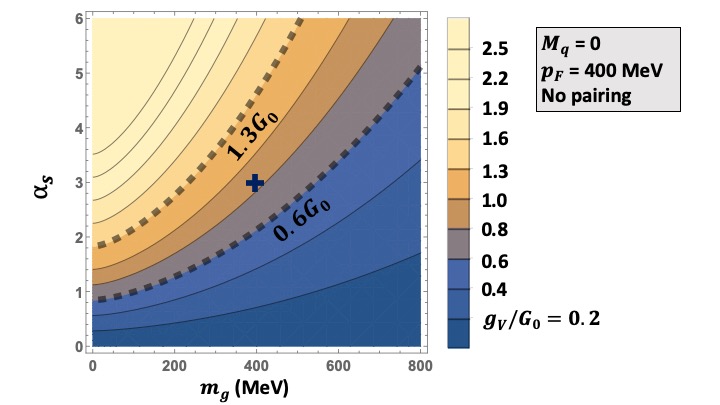}
\caption{\label{asmg}  (Color online) The vector coefficient $g_V$ generated by different constant $\as$ and gluon masses $m_g$, at $p_F = 400$ MeV ($\sim5n_0$).   The central cross indicates $\as$ = 3 and $m_g$= 400 MeV.}
\end{figure}

   Figure \ref{asmg} gives contour plots of the resulting vector coefficient $g_V$ for given different $\as$ and gluon mass $m_g$, at $5n_0$ and $20n_0$.    For the resulting $g_V/G_0$ to be in the interval 0.6-1.3 at $5n_0$ with $m_g = 400$ MeV, one needs a strong $\as \sim$ 2-4, 
   within the range of possible quark-gluon coupling strengths at low energies \cite{running}.
   Future theories of the quark-gluon vertex $\as$ together with detailed forms of gluon correlation functions below one GeV will be of interest as they can be directly related to effective quark models constrained by neutron star observations.  
   
   In the density range $\sim 5 n_0$ in a neutron star, where the quark Fermi momentum lies well below one GeV, it is reasonable to assume an approximately constant $\as$ and $m_g$.   The two limiting results, Eq.~(\ref{eq: g-V-limits}),  thus suggest an approximate density-dependent parametrization of $g_V$ based on explicit single-gluon exchange 
\beq
    g_V(p_F; m_g) \simeq \frac{4\pi \as/3}{9m_g^2 + 8 p_F^2}.
    \label{eq24}
\eeq
This parametrization is useful for including the density dependence of $g_V$ in the quark-hadron crossover equations of state.

\section{Effect of finite quark mass  \label{chiral}} 

        At high densities quark matter contains both a weak chiral condensate, $\sim \langle \bar q q\rangle$ as well as a diquark condensate $\sim \langle q q\rangle$, as a consequence of the six-quark Kobayashi-Maskawa-'t Hooft (KMT) effective interaction \cite{chiral1}.  The quark effective mass, $M_q \sim \langle \bar q q\rangle$, is  dynamically generated by the chiral condensate; in the NJL model, $M_q$ is the mean-field self-energy generated by the effective local four-quark interaction.  At densities $\gtrsim 5n_0$, the chiral condensate enhanced by the KMT interaction could result in an effective mass $M_q\sim 50$-$70$ MeV for the light quarks, and $\sim 250$-$300$ MeV for the $s$ quark \cite{NSReview}.  These masses are not small compared to the quark Fermi momentum at these densities, and must be taken into account in the exchange energy calculation.  
        
        Here we calculate the effects of $M_q$ on $g_V$ only by modifying the quark propagators in Eq.~(\ref{Egv}), and not further correcting the vertices.   We recognize that this is not a self-consistent calculation; rather we aim here to get a sense of the effects of a finite quark mass on the the vector channel of the matrix element (\ref{Eex}), which is connected to perturbative QCD at asymptotic density.   We take the quark Green's function to be
\begin{align}
S^{ab}_{ij}(p) & =  \delta_{ab}\delta_{ij} \frac{\gamma_\mu p^\mu +M_q}{(p_0+\mu)^2 - {\bm p}^2 -M_q^2}, 
\label{eq:S_Mq}
\end{align}
and assume the same effective mass $M_q$ for all flavors.

With this $S$, we obtain after some algebra, with $\epsilon_p = (|{\bf p}|^2 + M_q^2)^{1/2}$,
\begin{widetext}
\begin{align}
E_{\rm QCD}^{\rm v} & =  24\pi \alpha_s \left[  \left(\int \frac{d^{3}p}{(2\pi)^{3}}\ \frac{f(\epsilon_{p}-\mu)}{\epsilon_{p}}\right)^{2} - (2M_q^2-m_g^2) \int\frac{d^{3}p\,d^3p'}{(2\pi)^{6}}\frac{1}{\epsilon_{p}\epsilon_{p'}}\cdot\frac{f(\epsilon_{p}-\mu_q)f(\epsilon_{p'}-\mu_q)}{(\epsilon_{p}-\epsilon_{p'})^{2}-|{\bm p}-{\bm p'}|^2-m_g^2}     \right]. \nonumber \\
\label{eq:Eex_Mq}
\end{align}
\end{widetext}

\begin{figure}
\includegraphics[width=0.9\linewidth]{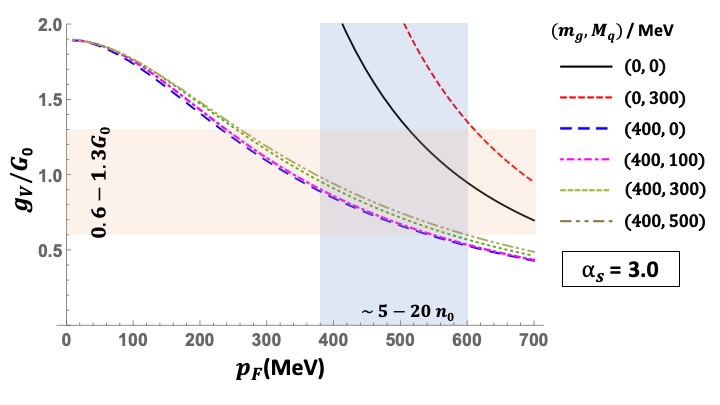}
\caption{\label{gvmqmg} (Color online) Vector repulsion coefficient $g_V$ for different values of  $M_q$ with $m_g=400$ MeV and
 $\alpha_s = 3$. }
\end{figure}

The asymptotic forms of Eq.~(\ref{eq:Eex_Mq}) for $p_F \gg M_q$ and $m_g$, and for $p_F \ll M_q$ and $m_g$ can be 
readily found, 
with the result that $g_{_V}(p_F; m_g, M_q)$  agrees in these limits with Eq.~(\ref{eq: g-V-limits}).
 In particular, $g_V$ is independent of $M_q$ at $p_F=0$ as long as $m_g$ is finite. 
 The combined effects of $M_q$ and $m_g$ are shown  in Fig.~\ref{gvmqmg}, which compares $g_{_V}$ at several different 
  values of $M_q$ and $m_g = 400$ MeV.  We find that 
   the effect of $M_q$ on $g_{_V}$ is almost negligible.
 Thus the assumption that $g_{_V}$ is flavor independent is reasonable, despite flavor symmetry being significantly broken by the strange quark mass; the parametrization (\ref{eq24}) is approximately useful independent of flavor.

\section{Effect of the diquark pairing \label{pair}}

We  next consider the effects on $E_{\rm QCD}^{\rm v}$ of scalar color-flavor-locked pairing among quarks through modification of the  normal  quark Green's function $S$ in Eq.~(\ref{Egv}).\footnote{The anomalous Green's function,
$F^{ab}_{ij}(x-y) = -i \langle \mathcal{T} q^a_i(x) (q^TC)^j_b(y) \rangle $, leads as well to the familiar energy shift $E_{\rm QCD}^{\rm pair}$ proportional to the square of the pairing gap, an effect  related to inferring the in-medium modification of $H$.}
In the CFL phase it is convenient to expand the quark field  (with SU(3) flavor and SU(3) color indices), as $q_{ia} = \sum_{A=0}^8  \lambda_{ia}^A q_A/\sqrt2$,  in term of the  Gell-Mann matrices, $\lambda^A$ ($A=1,2,...,8$), and  $\lambda^0 = \bm{1} \sqrt{2/3}$.
In this basis, the normal quark propagator becomes diagonal
\beq
S_{ij}^{ab}(x-y) = \sum_{A} \frac{1}{2} \lambda_{ia}^A \lambda_{bj}^A S_{A}(x-y).
\eeq
With CFL pairing, the $S_{A=1,...,8}$ describe eight paired quark quasiparticles  with the same gap $\Delta_{A=1,...,8}(p) = \Delta(p)$, and one quasiparticle $S_0$ with double the gap $\Delta_0(p) = 2\Delta(p)$.

  For  massless quarks ($\epsilon_p = |{\bf p}|$), one finds
\begin{widetext}
\beq
E_{\rm QCD}^{\rm v} &=&\frac{4\pi\as}{27}\sum_{A,B} \int_{pp'} \mbox{tr} [S_A(p)\gamma^\mu]  \mbox{tr} [S_B(p')\gamma_\mu] \frac{1}{(p-p')^2-m_g^2}, \\ 
 &= &  \frac{\as}{54\pi^3} \sum_{A,B} \int_0^{\infty} dp\,dp'\,v_{Ap}^2 v_{Bp'}^2 
\left[ 4pp' - J_{AB}(p,p',m_g) \ln \left| 1 + \frac{4pp'}{ J_{AB}(p,p',m_g)}\right| \right],
\label{eq49}
\eeq
\end{widetext}
 with $v_{Ap}^2  =  \frac{1}{2} \left(1- ({\epsilon_p - \mu })/{E_{p}^A } \right)$,
$E_p^A = [(\epsilon_p - \mu)^2 + \Delta_A^2]^{1/2} $, and 
$J_{AB}(p,p',m_g) = m_g^2 + (p-p')^2 - (E_p^A-E_{p'}^B)^2 $.
Generalization to the case with finite quark mass $M_q$ is straightforward.
Note that the total quark density is given by 
\beq
n_q = 2\sum_A \int \frac{d^3 p}{(2\pi)^3} v_{Ap}^2.
\label{nq}
\eeq

\begin{figure}
\includegraphics[width=1.1\linewidth]{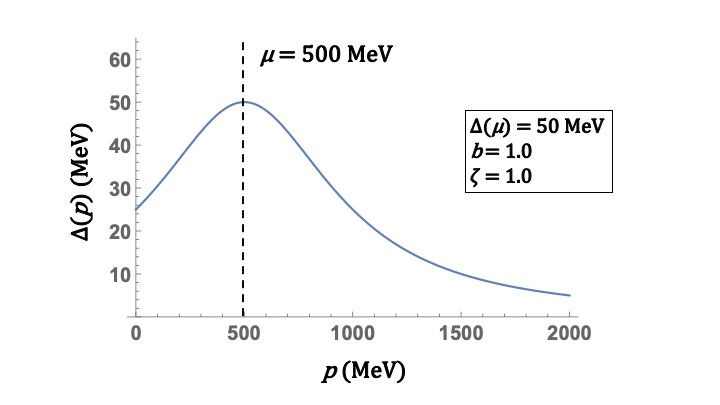}
\caption{\label{fig:gap}  The parametrization (\ref{eq37}) of the momentum dependence gap $\Delta(p)$ for $\mu = 500$ MeV, $b = 1.0$, $\Delta(\mu) = 50$ MeV, and $\zeta$ = 1.0.} 
\end{figure}

    The integral in Eq.~(\ref{eq49}) converges only with a momentum dependent gap. Following the numerical study in Ref.~\cite{spatial,Abuki2002}, we approximate the spatial momentum dependence of $\Delta$ by 
\beq
\Delta(p) = \frac{\Delta(\mu)}{(1 + b(p-\mu)^2/\mu^2)^\zeta} ;
\label{eq37}
\eeq
the constant $b>0$ parametrizes how fast $\Delta(p)$ falls off away from the Fermi surface, and the exponent $\zeta>0$ parametrizes the behavior of $\Delta(p)$ at high momenta (see Fig.~\ref{fig:gap}).  In the weak coupling limit, 
$\zeta = 1 + \mathcal{O}(\alpha_s)$ and  $\Delta \sim \mu g^{-5} e^{-3\pi^2/\sqrt{2}g}$ \cite{Son1999,Pisarski:1999tv}.
 Here we simply vary the gap in the range, $\Delta(\mu) = 100$-$300$ MeV, consistent with the QHC19 equation of state.

\begin{figure}
\includegraphics[width=.9\linewidth]{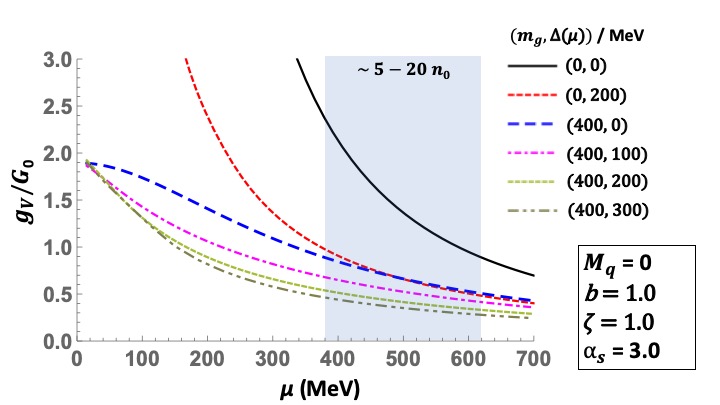}
\caption{\label{gvdmg}  (Color online) The vector repulsion coefficient $g_V$ for different  $\Delta(\mu)$ with $m_g=400$ MeV and $\alpha_s=3$.
The curves show how inclusion of pairing in the presence of a massive gluon has only a small effect on $g_V$.  } 
\end{figure}

 As we see, a gap decreases $g_V$ at all densities, and the dependence of the gap is significant for massless gluons. For gluon masses $m_g \sim 400$ MeV, however, even a large variation of $\Delta$ from $0$ to $300$ MeV does not change the qualitative behavior of  $g_V$.   In comparison with the effects of $M_q$,  a large gap $\Delta(\mu)=200$ MeV  (as in QHC19)  still has a sizable impact:  at $5n_0$, a 200 MeV CFL gap reduces $g_{V}$ from $\sim 0.9\, G_0$ to $\sim 0.55\, G_0$, even with $m_g = 400$ MeV.   

 The gluon propagator is also modified in a dense quark medium by Landau damping \cite{Pethick1989,Son1999,Pisarski:1999tv}, and the Debye screening mass in the longitudinal sector,  and in the presence of diquark pairing by Meissner masses in the transverse sector \cite{Fukushima2005, Rischke2000},
 of order $\sqrt\as \, \mu$.    The interplay of these modifications of the gluon propagator in the quark matter in neutron stars, and their effects on neutron star properties is an open question worthy of future research.


 \section{Conclusion \label{conclusion}}

  We have computed the vector repulsion coefficient $g_V$ from the explicit  gluon exchange energy in quark matter, modifying the quark and gluon Green's functions to account for a non-perturbative  gluon mass $m_g$, chiral condensate and diquark pairing, and included as well a possible infrared-finite $\as$.   In the density range $\sim 5$-$20 n_0$ with reasonable parameters for $\as$, gluon mass, quark mass and pairing gap, we can begin to understand the origin of a $g_V$ of order $\sim  0.6$-$1.3G$. The parameters we have chosen, despite their uncertainties, lie within estimates from a variety of models and theoretical frameworks of sub-GeV QCD.  Among the non-perturbative effects we have considered, the resulting $g_V$ is most sensitive to $\as$ and $m_g$, while $M_q$ and $\Delta$ induce only relatively small changes owing to suppression by a gluon mass.  Thus, the parametrization (\ref{eq24}) should be a good approximate description of the density dependence of $g_V$,  to be included in the equation of state for neutron star matter with a strongly interacting quark phase.
        
Many open questions remain.   The vector repulsion between quarks at densities $\gtrsim5n_0$ may also come from non-perturbative QCD beyond the single gluon-exchange contribution treated in this paper; such uncertainty is not under control at present.  
     As $\as$ could range anywhere from 0 to 10 (or even be divergent at low momentum scales), the assumption that the vector repulsion is dominated by a single gluon  exchange   with a fixed $\as$ and $m_g$ 
  is overly simplified.  Our treatment  can be improved and extended in several directions.  The first would be inclusion of more realistic quark and gluon propagators,  including possible momentum dependence of masses and differences between transverse and longitudinal gluons.  The second would be to include the non-perturbative running of $\as$.   Including the density dependence of $g_V$, as in the parametrization (\ref{eq24}), can have a significant effect on model studies of quark matter.  In particular, corrections to the contributions from the light and heavy quarks could shift the phase boundaries and modify the equation of state.   Including the density dependence of the diquark coupling, $H$, would have similar effect.
          
    We note that relating the effective QCD vector couplings $g_V$ and $g_\tau^{\alpha}$  (Appendix B) in the NJL model of dense matter (an effective field theory for quarks) to nucleon-meson models (effective field theories for hadrons)  would provide a further probe of quark-hadron continuity \cite{Schaefer1999,chiral1}.  If the transition from nuclear to quark matter is essentially smooth, one expects the vector repulsion from hadronic to quark matter to be similarly smooth, since in the quark-hadron continuity picture, the spectrum of light gluonic excitations is tightly connected to that of hadronic vector mesons  \cite{Hatsuda2008}, while quarks are mapped to the baryons in nuclear matter. Low energy quark-gluon matter treated in this way becomes an extension of the baryon-meson picture of nuclear matter, plausibly enabling a relatively smooth crossover and in turn mapping $g_V$ and $g_\tau^\alpha$ from the hadronic to quark phases.\footnote{ One may ask how vector repulsions in the nucleon-meson description of nuclear matter, a gauge-invariant theory, can be mapped onto vector repulsions in the gauge-dependent theory of quarks and gluons, despite the vector repulsions in both being effective fermion-fermion interactions mediated by massive boson exchange.  In fact, including color charge screening by CFL diquark condensates \cite{schafer2004, Yifan2019} leads to a low energy gauge-invariant description of quarks and gluons of the same form as a baryon-meson Lagrangian.  \cite{schafer2004}.}

\section*{Acknowledgments}
                Authors G. Baym and Y. Song are grateful to the RIKEN iTHEMS program for hospitality during this work.   Their research was supported in part by National Science Foundation Grant No. PHY1714042.  
Author T. Hatsuda was partially supported by the RIKEN iTHEMS program and JSPS Grant-in-Aid for Scientific Research (S), No. 18H05236, and Author T. Kojo by NSFC grant 11650110435 and 11875144,  and by the KMI for his long-term stay at the Nagoya University.
  The authors are grateful to the Aspen Center for Physics, supported by NSF Grant PHY1607611, where part of this research began, and to Hajime Togashi and Shun Furusawa for discussions there.

\appendix

\section{Fierz transformation \label{sec:fierz}}

The Fierz transformation is a re-arrangement of fermion operator products in the Dirac, flavor and color space using index-exchanging properties of the gamma and $SU(N)$ generator matrices.
In the quark-antiquark channel, re-arrangement of  the Dirac indices read
\beq
(\gamma^\mu)_{mn} (\gamma_\mu)_{m'n'} &=& \bm{1}_{mn'} \bm{1}_{m'n} + (i\gamma_5)_{mn'}(i\gamma_5)_{mn'} \nonumber \\
&&- \frac12 (\gamma^\mu)_{mn'} (\gamma_\mu)_{m'n} \nonumber \\
&& - \frac12(\gamma^\mu\gamma_5)_{mn'}(\gamma_\mu \gamma_5)_{m'n},
\eeq 
and those of the the flavor and color indices ($N_f = N_c = 3$) read 
\beq
\bm{1}_{ij}\bm{1}_{kl} = \frac13 \bm{1}_{il}\bm{1}_{kj} + \frac12 (\tau_a)_{il} (\tau_a)_{kj},\nonumber \\
\lambda^{ab}_\alpha \lambda^{a'b'}_\alpha = \frac{16}{9} \bm{1}_{ab'}\bm{1}_{a'b} - \frac13 \lambda^{ab'}_\alpha \lambda^{a'b}_\alpha.
\eeq         
In the quark-quark channel, 
\beq
(\gamma^\mu)_{mn} (\gamma_\mu)_{m'n'} &=& (i\gamma^5C)_{mm'}(i\gamma^5C)_{nn'} + C_{mm'}C_{nn'} \nonumber \\
&& -\frac12 (\gamma^\mu \gamma^5C)_{mm'}(\gamma_\mu \gamma^5C)_{nn'} \nonumber \\
&& - \frac12 (\gamma^\mu C)_{mm'}(\gamma_\mu C)_{nn'} ,
\eeq    
and
\beq
\bm{1}_{ij}\bm{1}_{kl} &=& \frac12(\tau_S)_{ik} (\tau_S)_{lj} + \frac12(\tau_A)_{ik} (\tau_A)_{lj}, \nonumber \\
\lambda^{ab}_\alpha \lambda^{cd}_\alpha &=& \frac23 \lambda^S_{ac} \lambda^S_{bd} - \frac43 \lambda^A_{ac} \lambda^A_{bd},
\eeq     
where $S$ and $A$ stand for symmetric and antisymmetric indices, and the $\tau_{\alpha = 1,\dots,8}$ are the eight Gell-Mann flavor matrices.   
 Using these relations, one can transform a single trace into products of two traces, as done in e.g. Eq.~(\ref{Egv}):
\beq
\Tr [S(p) \Gamma^I S(p') \Gamma^I] = \sum_M g_M \Tr [S(p) \Gamma^M] \Tr[S(p') \Gamma^M], \label{A2} \nonumber \\
\eeq
where $\Gamma^I$ are Dirac, flavor and color matrices.

\section{The vector-isovector interaction \label{sec:nucleon}}

 The discussion in the main body of the text focusses on the flavor symmetric case, where in the absence of pairing  the vector component of single gluon exchange contributes only to the isoscalar channel.    (In the CFL phase, one finds non-vanishing contributions in the flavor-color vector channel $(\bar q \gamma^\mu \tau_a \lambda_A q)^2$ as well.)    For realistic constituent quark masses, however, the vector-isovector channel (denoted by $\tau$), corresponding to the interaction 
$(\bar q \gamma^\mu \tau_\alpha q)^2$, also contributes to the 
single gluon exchange energy,  
\beq
E_{\rm QCD}^{\rm v, \tau} &=& 
 \frac{2\pi \as}{9} \int_{p,p'} \Tr [S(p)\gamma^\mu \tau_\alpha ] \Tr [S(p')\gamma_\mu \tau_\alpha] D(p-p').\nonumber \\
\eeq
 In particular, the $\alpha$ = 3 and 8 terms yield 
 the exchange energy at low density of the form,
\beq
g_\tau^{(3)}  (n_u-n_d)^2 + \frac{g_\tau^{(8)}}{3} (n_u+n_d-2n_s)^2 .
\label{gi}
\eeq 
This vector-isovector energy is analogous to the neutron-proton symmetry energy in nuclear matter.  For single gluon exchange, 
$g_\tau^{(3)} = g_\tau^{(8)} = \frac32 g_V$, indicating an vector-isovector energy comparable to the vector-isoscalar energy for significant differences in flavor densities. It is an interesting future problem to estimate the in-medium values of 
$g_\tau^{(3,8)}$ as well as $g_V$  by matching with,  e.g.,  the chiral nucleon-meson model \cite{weise}.

\section{Estimating $H$ from the $N-\Delta$ mass splitting \label{sec:Ndelta} }

Another important ingredient in the QHC19 equation of state is the parameter $H$ that quantifies the strength of attractive diquark correlations.   At high density diquark correlations are the driving force of color superconductivity, while at low density the correlations appear in the context of hadron mass splittings, e.g., the $N$-$\Delta$ splitting, $m_\Delta -m_N \simeq 293$ MeV. The density $n_B \sim 5n_0 \simeq 0.8\,{\rm fm}^{-3}$ is roughly that inside of baryons, and so suggests the possibility of inferring the value of $H$ at $n_B \sim 5n_0$ from the $N$-$\Delta$ splitting.   

  This splitting has been derived by Ishii et al.~\cite{Ishii}, by solving the Faddeev equations of three-quark systems within the NJL model.   They included effective four-quark interactions in the isoscalar scalar and isovector axial-vector diquark channels, which in our notation are:
\begin{eqnarray}
{\cal L}_S 
&=& H \sum_{A=2,5,7}  \left( \bar{\psi} i \gamma_5 \tau_2 \lambda_A \psi_C \right)  \left( \bar{\psi}_C i \gamma_5 \tau_2 \lambda_A \psi \right),
 \\
{\cal L}_A 
&=& H' \! \sum_{A=2,5,7}  \left( \bar{\psi}  \gamma_\mu  \tau_2 \vec{\tau}  \lambda_A \psi_C \right)  \left( \bar{\psi}_C  \gamma^\mu \tau_2 \vec{\tau} \lambda_A \psi \right).
\end{eqnarray}
Reference~\cite{Ishii} finds the approximate formulae
\begin{eqnarray}
M_N &\simeq& 1.70 - 0.21 r'_H - 0.33 r_H  ~~~[{\rm GeV}]\,, \\
M_\Delta &\simeq& 1.52 - 0.22 r'_H ~~~[{\rm GeV}]\,.
\end{eqnarray}
where $r_H = H/G_0$ and $r_H' = H'/G_0$.
The absolute values of these masses are not quite trustworthy as they are sensitive to the physics beyond the NJL model, e.g., confinement.  In the mass splitting such uncertainties are largely cancelled and the physics of short-range correlations become dominant.  Using the empirical $M_\Delta - M_N$  we find
\beq
- 0.01 r'_H + 0.33 r_H  \simeq 0.47~~~[{\rm GeV}] \,.
\eeq
Provided $r'_H \ge 0$ as expected from typical models, we arrive at
\beq
H/G_0 \gtrsim 1.4,
\eeq
consistent with the range in QHC19, $H/G_0$ =1.35 -1.65.  More comprehensive studies will be given elsewhere \cite{H_2019}.

\end{document}